\documentclass[aip,apl,reprint,graphicx]{revtex4-2}

\usepackage[pdftex]{graphicx}
\usepackage{amsmath}
\usepackage{verbatim}
\usepackage{xcolor}
\usepackage{bm}
\usepackage{txfonts}
\usepackage{siunitx}
\usepackage{float}
\usepackage[colorlinks=true,allcolors=blue]{hyperref}
\usepackage[utf8]{inputenc}
\usepackage[T1]{fontenc}
\usepackage{soul}

\draft

\newcommand{\changed}[1]{\textcolor{black}{#1}}

\begin{document}
\title{Bismuth-substituted Lutetium Iron Garnet Films with Giant Visible-Range Magneto-Optical Sensitivity}
\author{Megan H.\ Dransfield}
\author{Lukáš Flajšman}
\email{lukas.flajsman@aalto.fi}
\author{Matthijs H.\ J.\ de Jong}
\author{\changed{Lars Peeters}}
\author{\changed{Rhodri Mansell}}
\author{Sebastiaan van Dijken}
\author{Laure Mercier de Lépinay}
 \email{laure.mercierdelepinay@aalto.fi}
\affiliation{Department of Applied Physics, Aalto University, FI-00076, Finland}

\date{\today}

\begin{abstract}

    Magneto-optical materials are indispensable across modern physics, serving as the foundation for precision magnetic sensing, nonreciprocal photonics, and optical isolation technologies. The continual pursuit of materials with high Verdet constants has driven the development of garnet-based compounds exhibiting \changed{giant} magneto-optical sensitivity. In this work, we report the growth and comprehensive magneto-optical characterization of bismuth-substituted lutetium iron garnet (LuBiIG), a material that combines the large spin-orbit coupling of bismuth with the lattice stability of lutetium iron garnet. LuBiIG films with thicknesses between \changed{\SIrange{85}{220}{\nano\metre}} were grown by pulsed laser deposition and characterized at room temperature over the \SIrange{500}{820}{\nano\metre} wavelength range. The films exhibit an exceptionally high Verdet constant of up to \changed{\SI{-1.37e8}{\degree\per\metre\per\tesla}}, peaking in the visible spectral range near \SI{520}{\nano\metre}. These results position LuBiIG as a highly sensitive magneto-optical material suitable for advanced cryogenic detection and hybrid quantum applications.
\end{abstract}
\maketitle

%\section{Introduction}

The development of highly sensitive magneto-optical materials is essential for imaging and detecting nanoscale magnetic textures and weak magnetic fields, including magnetic domains, skyrmions, and flux structures in superconductors~\cite{Simpson2016,Wells2016,Goa2003}. Beyond fundamental studies, such materials underpin technological applications ranging from high-density magnetic recording~\cite{Dorosinskiy2023,Tsunashima2001,Kryder1985} to optical isolators and nonreciprocal photonic components~\cite{Srinivasan2022,Jalas2013}. Over the past decades, extensive research has explored %various 
rare-earth and transition-metal garnets \changed{to enhance} available magneto-optical sensitivity%and spectral tunability. 
. In magneto-optical detection, linearly polarized light interacts with a magnetized transparent medium, and its polarization is rotated by the Faraday or Kerr effect. The magnitude of this rotation directly encodes the local magnetic field distribution, providing a powerful, non-invasive means to map out magnetic textures. \changed{When used for magnetic field imaging, such films are referred to as magneto-optical indicator films \cite{Path2024,Goa2003,Wells2016}. Our goal is to realize such thin films optimized for quantitative magneto-optical imaging.}

Substituted or doped garnet films combine strong magneto-optical coupling, high transparency, and large magnetic permeability~\cite{Nakamura2024}\changed{, making them} uniquely suited for high-resolution optical magnetometry and \changed{indicator films}. Numerous studies have shown that the magneto-optical performance of garnets can be greatly enhanced by rare-earth or heavy-element substitution~\cite{Shinagawa1998,Hansen1986,Helseth2001}. 
\changed{Bismuth is a particularly promising dopant, as its strong spin-orbit coupling increases the Faraday rotation almost linearly with concentration \cite{Hansen1985a}}. Yet, the fabrication of bismuth-substituted iron garnets remains challenging due to the volatility of bismuth and its sensitivity to growth conditions. To produce such materials, most previous studies have relied on liquid-phase epitaxy (LPE). \changed{This technique} yields excellent crystalline quality but often leads to film cracking and non-uniform bismuth incorporation~\cite{Johansen1976}, both of which degrade magneto-optical performance. To overcome these limitations, pulsed laser deposition (PLD) can be used as a versatile, nonequilibrium technique capable of preserving target stoichiometry. \changed{It also} enables precise control over oxygen pressure and growth rate. This approach maintains the desired bismuth concentration while ensuring smooth, crack-free epitaxial films.

Bismuth-lutetium iron garnets (LuBiIG) have been recognized for several decades for their exceptionally high Verdet constants, and reduced absorption in the near-infrared range~\cite{Hansen1985a,Hansen1984,Rojas2004}. \changed{Such films have demonstrated} Faraday rotation reaching \SI{4.8e6}{\degree\per\metre} at \SI{633}{\nano\metre}\cite{Hansen1984a}. \changed{Here, using Bi$_{2.1}$Lu$_{0.9}$Fe$_5$O$_{12}$ films,} we report Faraday rotation as high as \changed{\SI{6.3e6}{\degree\per\metre} at \SI{630}{\nano\metre}}. The enhancement \changed{of the Faraday effect compared to that of YIG or LuIG} originates from \changed{the strong spin-orbit coupling of Bi and the hybridization of Bi and Fe-O orbitals, which lift the degeneracy of the electronic transitions involved in magneto-optical response of garnets \cite{Deb2019}}. %originates from diamagnetic electronic transitions between the ground and triply degenerate excited states which strongly modify the permittivity tensor, as well as from crystal-field-induced transitions within the Fe sublattices~\cite{Silliman1992}. 
Lutetium substitution alleviates the lattice mismatch between Bi:Fe garnets and gadolinium gallium garnet\changed{-based} (GGG) substrates, stabilizing epitaxial growth and ensuring stoichiometric fidelity~\cite{Hansen1985}. Additionally, these materials have demonstrated sensitivity down to cryogenic temperatures (\SI{4}{\kelvin}) \cite{Veshchunov2016}, \changed{with increasing Faraday rotation as the temperature is decreased \cite{Hansen1985a}, making} LuBiIG a prime candidate for high-sensitivity magneto-optical imaging at low temperatures.

This work \changed{reports on the PLD growth of} LuBiIG films suited for the imaging of nanoscopic superconducting vortices hosting quantized magnetic flux that penetrates type-II superconductors. For this application, the films would be mounted in a flip-chip stack with superconducting samples allowing for an approach within \SIrange{200}{1000}{\nano\metre} of magnetic structures. To detect stray vortex fields of the order of \SI{1}{\milli\tesla} or less, the magneto-optical sensor must be highly sensitive. Moreover, to resolve single vortices, the film thickness should ideally be comparable to the magnetic-field attenuation length of a vortex, typically around \SI{100}{\nano\metre} for superconductors in the thin film limit.
\changed{This paper reports the growth and room-temperature characterization of crystalline LuBiIG films of thickness \SIrange{85}{220}{\nano\metre} with ultra-high Verdet constants. Using the room temperature results as a conservative estimate for low temperature performance, we discuss the feasibility of realizing cryogenic magneto-optical microscopy of vortex textures.}

In preparation for film growth, (111)-oriented \changed{substituted-}GGG substrates \changed{substituted-}GGG substrates \changed{(sGGG, in which Ca, Mg and Zr partially substitute for Gd and Ga)} ($5\times5\times0.5$~\si{\cubic\milli\metre}) were ultrasonically cleaned in acetone and isopropanol and pre-annealed at \SI{1000}{\celsius} for four hours in flowing oxygen. \changed{This treatment significantly improves the substrate surface quality, promoting reproducible epitaxial growth.} \changed{The substituted-GGG (sGGG) substrates have a lattice parameter of \SI{1.249}{\nano\metre}, larger than ordinary GGG substrates. This reduces the lattice mismatch to the Bi$_{2.1}$Lu$_{0.9}$Fe$_5$O$_{12}$ layer, which is expected to have a lattice constant around \SI{1.250}{\nano\metre} \cite{Hansen1985a}}.

Targets of composition Bi$_{2.1}$Lu$_{0.9}$Fe$_5$O$_{12}$ were employed in a commercial PLD system. PLD parameters have been previously explored for \changed{LuBiIG/GGG films} \cite{Leitenmeier2006}, although with \changed{a much lower Bi} concentration. Films (see Table~\ref{table:1} for the complete list) were deposited using a KrF excimer laser (\SI{248}{\nano\metre}) operating at a fluence of \SI{2.5}{\joule\per\centi\metre\squared} and a repetition rate of \SI{5}{\hertz}. The substrate temperature during deposition was systematically varied between 650 and 850$\,\rm ^\circ C$, while the oxygen background pressure ranged from \SIrange{0.025}{0.15}{\milli\bar}. Each film was obtained by 15,000--25,000 laser pulses, yielding thicknesses in the range of \changed{\SIrange{85}{220}{\nano\metre}}. 

\changed{Following deposition, the samples were cooled \textit{in situ}, either at the growth pressure or after post-annealing in \SI{100}{\milli\bar} of oxygen for \SI{30}{\minute}.} The annealing atmosphere was selected based on reflection high-energy electron diffraction (RHEED) \changed{images obtained after the growth}. \changed{Films displaying sharp RHEED patterns, whether streak-like two-dimensional (2D) or spot-like three-dimensional (3D), were cooled at the deposition pressure, while those showing diffuse diffraction features were post-annealed in \SI{100}{\milli\bar} to promote crystallization and surface smoothening.} All samples were subsequently cooled to room temperature at a controlled rate of \SI{5}{\celsius\per\minute}. \changed{The post-growth RHEED patterns are shown in Fig.~S1 of the Supplementary Material}.

\changed{Energy-dispersive X-ray spectroscopy (EDS) carried out on samples 1, 5 and 8 confirms that the films reproduce the nominal composition of the target, within experimental error. The measured cation ratios for representative films are reported in Table~S1 of the Supplementary Material.} \changed{Additionally, the transmittance spectra for each sample was measured, and is displayed in Fig.~S2 in the Supplementary material. Overall, the transmittance increases as the wavelength approaches the near-infrared range. Jumps in the transmittance indicate electronic transitions at around \SI{450}{\nano\metre} and \SI{520}{\nano\metre}, consistent with reports on similar materials \cite{Deb2019}.}

\begin{table}[h!]
\centering
\caption{Growth-related parameters for LuBiIG films grown by PLD on sGGG substrates in O$_2$ atmosphere, along with the observed post-growth RHEED diffraction signatures indicating the crystalline quality of each film. The thickness $t$ \changed{was determined by stylus profilometry for samples~1--3, and from Laue oscillation fitting for samples~4--8.}}
\label{table:1}
\begin{tabular}{ |c|c|c|c|c|c|c| } 
 \hline
 Sample & Growth T  & Growth P & Cooldown P & Pulses & RHEED & $t$ \\
  &(\si{\celsius}) & (\si{\milli\bar}) &(\si{\milli\bar}) & x1000& pattern& (\si{\nano\metre})\\
 \hline
 1 & 850 & 0.025 & 100 & 15 & 3D& 164\\ 
 \hline
 2 & 750 & 0.025 & 0.025 & 25 & 3D& 218\\ 
 \hline
 3 & 650 & 0.025 & 100 & 15 & 2D+3D& 86\\ 
 \hline
 4 & 850 & 0.1 & 0.1 & 15  & 2D& \changed{\num{99+-3}}\\ 
 \hline
 5 & 750 & 0.1 & 0.1 & 15 & 2D& \changed{\num{92+-3}}\\ 
 \hline
 6 & 650 & 0.1 & 0.1 & 15 & 2D+3D& \changed{\num{85+-7}} \\ 
 \hline
 7 & 750 & 0.15 & 0.15 & 15 & 2D& \changed{\num{95+-4}}\\ 
 \hline
 8 & 650 & 0.15 & 100  & 15 & 2D+3D& \changed{\num{94+-4}}\\ 
 \hline
\end{tabular}

\end{table}

\begin{figure}[t]
    \centering
    \includegraphics[width=\columnwidth]{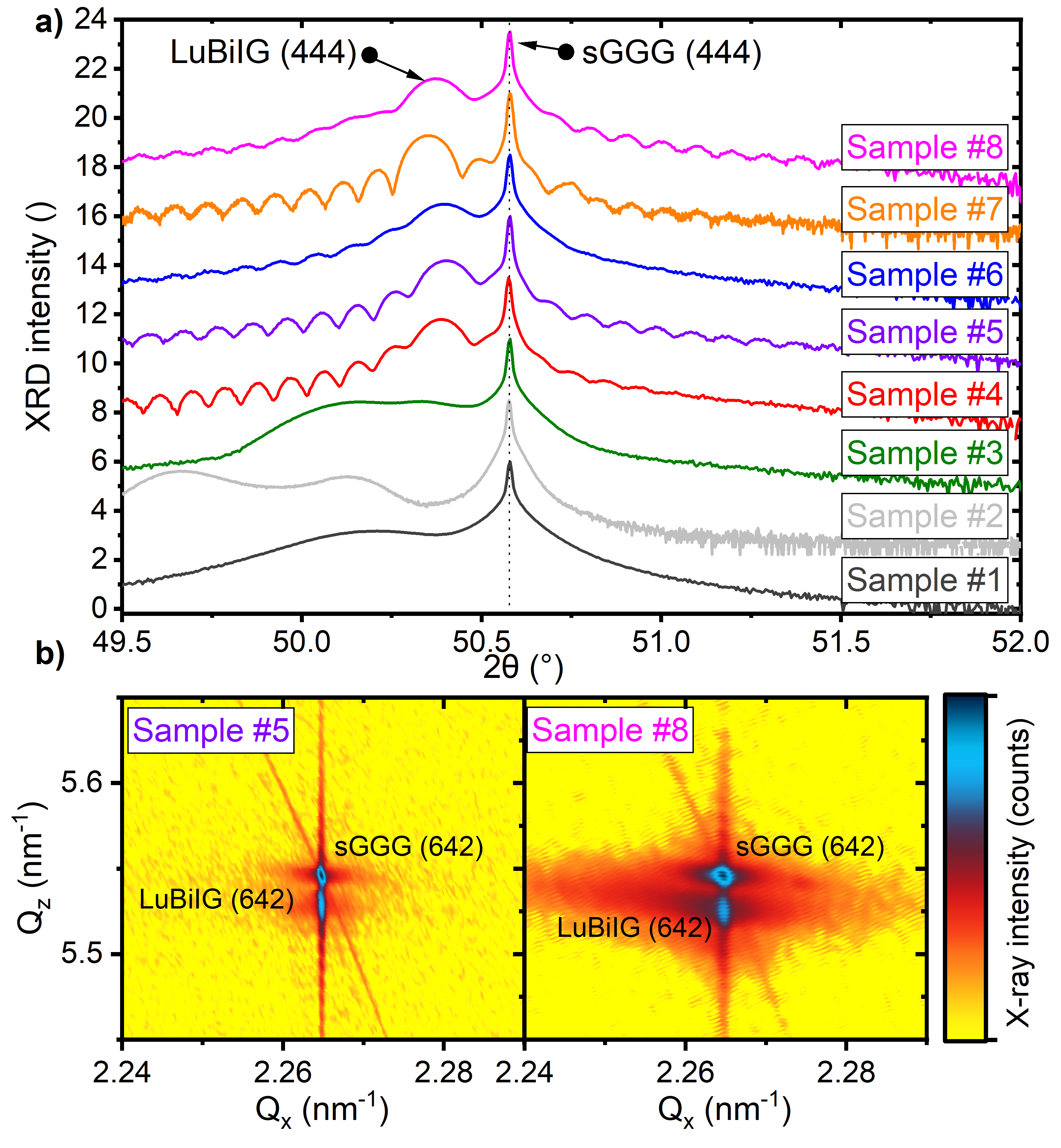}
    \caption{\changed{(a)~Symmetric $\theta$--$2\theta$ XRD scans around the (444) reflection for LuBiIG samples~1--8 (vertically offset for clarity). Samples~4--7 show a sharp LuBiIG~(444) film peak with Laue thickness fringes, indicating smooth, laterally uniform, high-quality epitaxial films; samples~1--3 show broad, multiphase features; samples~6 \& 8 shows a film peak with attenuated fringes. (b)~Reciprocal space maps around the asymmetric (642) reflection for sample~5 (left) and sample~8 (right). Both films share the same $Q_x$ as the sGGG substrate, confirming coherent, pseudomorphically strained growth. The reflection of sample~5 is sharp, whereas that of sample~8 is markedly broadened along $Q_x$, signalling mosaic spread.}}
    \label{fig:xrd}
\end{figure}

\changed{The crystalline quality and epitaxial relationship of the films were characterized by X-ray diffraction (XRD). Fig.~\ref{fig:xrd}(a) shows symmetric $\theta$--$2\theta$ scans recorded around the (444) reflection for the complete sample series. Samples~4--8 exhibit a well-defined LuBiIG~(444) film peak accompanied by Laue thickness oscillations, which attest to smooth surfaces, laterally uniform thickness, and a high degree of crystalline order~\cite{Hyun2025,Flajsman2025}. Samples~1--3, in contrast, display broad and multiple diffraction features indicative of mixed crystalline phases and poor epitaxial quality. Samples~6 and 8 have attenuated Laue oscillations, consistent with the partial structural disorder already inferred from its mixed 2D+3D RHEED pattern. No secondary-phase reflections are detected for the high-quality films, confirming single-phase garnet growth~\cite{Flajsman2025}. The LuBiIG~(444) film peaks are at lower angles than the sGGG~(444) peak indicating an expansion of the out-of-plane lattice parameter of the LuBiIG layer due to a compressive in-plane strain, which is estimated at around 0.2\%. Additionally, the relaxed LuBiIG lattice parameter is found to be around 1.252 nm (see Supplementary material section S5 for details).}

\changed{For samples~4--8 the thickness was determined by fitting the Laue thickness oscillations, while for samples~1--3} the film thicknesses were measured using a stylus profilometer (Bruker DektakXT). During deposition, a narrow section of each substrate was masked by a metal clamp, preventing film growth and thereby creating a sharp step edge that served as a zero-thickness reference. Thickness values extracted from these measurements are summarized in Table~\ref{table:1}. 

\changed{To probe the in-plane epitaxy, reciprocal space maps (RSMs) were acquired around the asymmetric (642) reflection (Fig.~\ref{fig:xrd}(b)). For sample~5, the LuBiIG~(642) reflection lies directly below the sGGG~(642) substrate peak at an identical in-plane scattering vector $Q_x$ and exhibits only a very small spread, demonstrating fully coherent, pseudomorphically strained heteroepitaxy in which the film adopts the in-plane lattice parameter of the substrate~\cite{Hyun2025,Flajsman2025}. For sample~8, by contrast, the film reflection is markedly broadened along $Q_x$. This indicates a distribution of crystallite orientations and in-plane lattice spacings, due to the onset of partial strain relaxation, in agreement with its broadened $\theta$--$2\theta$ peak and mixed RHEED signature.}% We note that this structural ranking does not map monotonically onto the magneto-optical response of Table~\ref{table:2}: the highly coherent sample~5 and the more mosaic sample~8 are both among the most magneto-optically active films, showing that the Verdet constant is not governed by crystalline perfection alone.}

\changed{The surface properties were further characterized by atomic force microscopy (AFM). The measured RMS roughness is as low as \SI{1.0}{\nano\metre} and \SI{1.5}{\nano\metre} for samples 5 and 7, respectively, where the measured roughness correlates with the XRD data shown above. Further details are given in Section~S4 of the Supplementary Material.}

%(b)~Reciprocal space maps around the asymmetric (642) reflection for sample~5 (left) and sample~8 (right). Both films share the same $Q_x$ as the sGGG substrate, confirming coherent, pseudomorphically strained growth. The sample~5 reflection is sharp, whereas that of sample~8 is markedly broadened along $Q_x$, signalling mosaic spread.

\changed{The magnetic properties of the LuBiIG films were characterized using longitudinal magneto-optical Kerr effect (MOKE) magnetometry in a nearly-crossed polarizer configuration. A continuous-wave \SI{405}{\nano\metre} laser was incident on the sample at an angle of \SI{45}{\degree}, and the reflected intensity was recorded as a function of the applied in-plane magnetic field. The plane of incidence and reflection was aligned along the in-plane crystallographic [1$\bar{1}$0] direction of the sGGG substrate. Hysteresis loops recorded for different in-plane field orientations were essentially indistinguishable, indicating that the films exhibit no significant in-plane magnetic anisotropy. In this longitudinal configuration the measured Kerr signal is proportional to the in-plane magnetization component along the applied field, so that the recorded loops directly probe the in-plane magnetization reversal.}

\begin{figure}[t]
    \centering
    \includegraphics[width=\columnwidth]{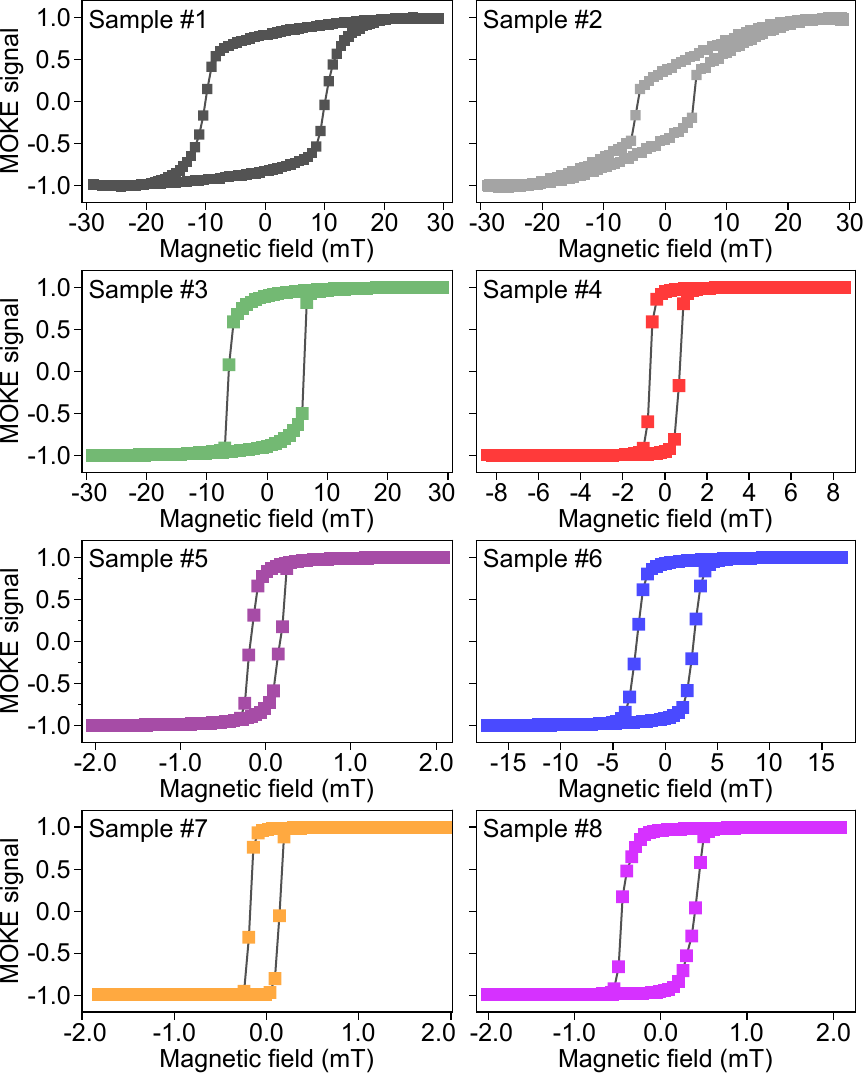}
    \caption{Longitudinal MOKE hysteresis loops for the series of LuBiIG samples. Each curve is normalized to its maximum intensity. Note that the magnetic field range (x-axis) differs between panels to highlight the variation in saturation behavior.}
    \label{fig:2}
\end{figure}

Magnetic hysteresis curves are shown in Fig.~\ref{fig:2}: all films exhibit low saturation fields in the in-plane direction. \changed{This indicates that the magnetization preferentially lies in the film plane, consistent with the compressive in-plane film strain~\cite{Hansen1985} seen in Fig.~\ref{fig:xrd}. The lowest coercive fields are found in the layers with the sharper LuBiIG~(444) film peaks and clearer Laue oscillations in XRD, as well as the lowest surface roughness in AFM. Poorer crystalline quality correlates with broader hysteresis loops due to an increase in magnetic disorder and surface roughness which enhance domain wall pinning \cite{Jeudy2026,Zhao2001}.} %The magnitude of the in-plane saturation and coercive fields varies notably across the sample series, reflecting differences in crystalline quality and oxygen stoichiometry. Samples showing sharp, streak-like 2D RHEED patterns indicative of regular layer-by-layer growth display narrow, square hysteresis loops with low coercive fields of a few millitesla. In contrast, films exhibiting mixed or diffuse 3D RHEED features show broader, slanted loops with higher coercivity and increased saturation fields, characteristic of enhanced pinning and structural disorder. \changed{Mechanistically, we envisage that the low pinning-site density of the well-ordered 2D-growth films allows magnetization to reverse by rapid domain nucleation and essentially unimpeded domain-wall propagation, producing square loops that switch within a few millitesla; the compressive epitaxial strain on sGGG(111) further sets the in-plane easy axis along which the field is applied. Conversely, the additional disorder and grain boundaries of the mixed-mode and 3D films pin domain walls, which broadens and slants the loops and raises both the coercivity and the in-plane saturation field.} These correlations confirm that optimized oxygen pressure and improved crystalline order yield magnetically softer LuBiIG films with coherent domain reversal conditions.

The saturation magnetization of each LuBiIG film was determined by vibrating sample magnetometry (VSM) under an in-plane magnetic field and the values are listed in Table~\ref{table:2}. \changed{The lowest saturation magnetizations (samples 1 \& 2) correlate with the disordered film structure shown by XRD, while the other films cluster around a $\mu_0 M_\mathrm{s}$ of \SI{170}{\milli\tesla}, similar to previous work \cite{Hansen1985a,Hansen1984}. }

%The measured saturation magnetization correlates approximately with the crystalline quality of the samples, assessed by RHEED: a high crystalline quality (evidenced by a pattern typical of 2D structures) produces a larger magnetization value.

The spectral magneto-optical response of the LuBiIG films was characterized using normal-incident linearly polarized light. A supercontinuum laser (NKT SuperK EXW-12) combined with an acousto-optic tunable filter provided a continuously tunable wavelength range from \SIrange{500}{1100}{\nano\metre}. The beam was passed through a high-extinction-ratio linear polarizer and directed onto the sample between the poles of an electromagnet, generating an out-of-plane magnetic field. Both the transmitted and reflected light were measured simultaneously to quantify the Faraday and Kerr effects.

\changed{To enable high-sensitivity lock-in detection, a photoelastic modulator (PEM) operating at \SI{50}{\kilo\hertz} was placed in the optical path to impose a polarization modulation. The transmitted beam was split into its orthogonal linear components using a polarizing beam splitter and recorded by independent photodiodes. Demodulation of the optical signal at the second harmonic frequency ($2\omega$) directly yielded the Faraday rotation amplitude\cite{Fernandez2020,Kataja2016}. This approach achieves sub-milliradian resolution across the visible-to-near-infrared spectral range under applied magnetic fields.}%Photoelastic modulators (PEMs) operating at \SI{50}{\kilo\hertz} were placed in the transmitted and reflected paths to impose a high-frequency polarization modulation. The resulting optical signals were demodulated using lock-in detection at the first and second harmonics of the modulation frequency. The first harmonic corresponds to the ellipticity component, while the second harmonic encodes the polarization rotation. The beams were subsequently separated into their orthogonal linear components using polarizing beam splitters, and each polarization was detected by an independent photodiode. Using the PEM reference signal as the lock-in amplifier input allowed precise extraction of both rotation and ellipticity amplitudes \cite{Fernandez2020,Kataja2016}. This approach provides high sensitivity, enabling detection of sub-milliradian polarization changes across the full visible-near-infrared range, and across a range of applied fields. \hl{Was there any data from the Kerr effect or the Faraday ellipticity? Why mention it here and never again?}
%Sweeping a calibrated out-of-plane magnetic field, the Verdet constant $V$ can be calculated from the polarization rotation $\theta$ as a function of the applied magnetic field $B$ using $V=\frac{1}{t} \frac{d \theta}{d B}$ where $t$ is the thickness of the film reported in table \ref{table:1}.
The setup was calibrated using a commercial magneto-optical sensor with a known magneto-optical response (Matesy sensor type B). \changed{The absolute magnitude of the magneto-optical rotation was further verified using a custom rotating-analyzer polarimeter (RAP) setup, exhibiting excellent agreement with the PEM-based calibration (see Section~S6 and Fig.~S4 in the Supplementary Material).}
%The absolute magnitude of the magneto-optical rotation was independently verified using a rotating-analyzer polarimeter (RAP) equipped with a motorized analyzer stage. For each magnetic field step, the analyzer was rotated around the extinction position within a \(\pm\SI{5}{\degree}\) angular range. The resulting transmission intensity was recorded as a function of analyzer angle, and the position of the minimum in each trace was used to determine the sample-induced polarization rotation. The values obtained from the RAP measurements showed excellent agreement with those acquired using the PEM-based detection system, confirming the accuracy of the magneto-optical calibration. \hl{This could go in the supplementary if you were also supplying some data, otherwise I don't see why this is adding anything.} \newline

%\subsection{Magneto-optical characterization}

\begin{figure*}
    \includegraphics[width=\linewidth]{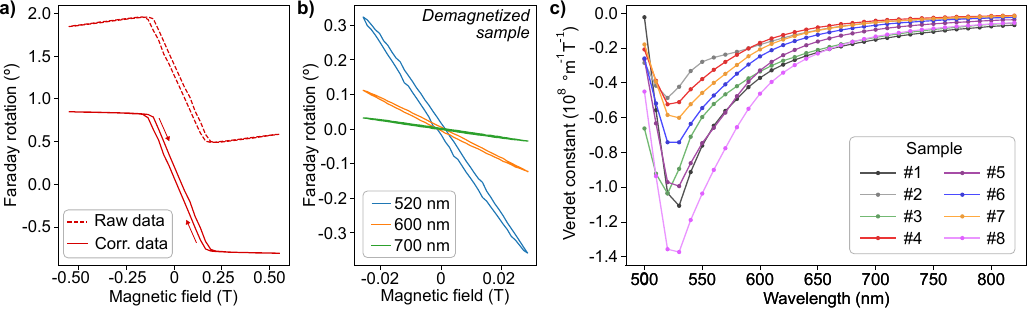}
    \caption{(a) Representative Faraday hysteresis loop of sample~8 measured at a wavelength of \SI{600}{\nano\metre}. The contribution from the sGGG substrate is subtracted from the raw data (dashed red line), and the linear regions of the corrected curve (solid red line) are fitted to extract the saturation field. (b)  Repeated hysteresis measurements at several wavelengths on a demagnetized sample over a reduced field range. The fitted slope is used to determine the Verdet constant. (c) Wavelength dependence of the Verdet constant for all LuBiIG films, exhibiting a pronounced maximum in absolute value at \SIrange{520}{530}{\nano\metre}.  }
    \label{results}
\end{figure*}

Fig.~\ref{results}(a) shows a representative Faraday rotation hysteresis loop of sample~8 measured at a wavelength of \SI{600}{\nano\metre}. The raw signal includes contributions from both the LuBiIG film and the sGGG substrate, the latter exhibiting a weak, linear magneto-optical response arising from its paramagnetic behaviour. \changed{To isolate the intrinsic film response, the gradient attributed to the linear sGGG response was subtracted from the data, along with} the small zero-field offset attributed to instrumental artifacts. 
The out-of-plane Faraday rotation curves of all samples exhibit a nearly linear approach to saturation with narrow hysteresis and much higher saturation fields than under in-plane fields consistent with a hard-axis magnetic response.
 Linear fits of the \changed{slope} and saturation plateaus were used to determine the saturation fields and the maximum Faraday rotation. The saturation fields averaged over all measured wavelengths are summarized in Table~\ref{table:2}, with uncertainties corresponding to the standard deviation of the wavelength-dependent fits.

\begin{comment}

\begin{table}[h]
\centering
\begin{tabular}{ |c|c|c|c|c| } 
 \hline
 Sample & OOP sat. field & Max Verdet const. & Max rotation & $\mu_0 M_\mathrm{s}$ \\ 
 & (\si{\milli\tesla}) & (\si{\degree/\micro\meter/\milli\tesla}) & (\si{\degree}) & (\si{\milli\tesla})\\
 \hline
 1 & 150 $\pm$ 54 & -0.109 & 2.96 & 228\\ 
 \hline
 2 & 206 $\pm$ 5 & -0.048 & 4.37 & 330\\ 
 \hline
 3 & 234 $\pm$ 4 & -0.103 & 2.55 & 662\\ 
 \hline
 4 & 347 $\pm$ 7 & -0.062 & 2.13 & 785\\ 
 \hline
 5 & 229 $\pm$ 9 & -0.106 & 2.36 & 537\\ 
 \hline
 6 & 304 $\pm$ 13 & -0.067 & 2.23 & 597\\ 
 \hline
 7 & 349 $\pm$ 7 & -0.062 & 2.33 & 1100\\ 
 \hline
 8 & 180 $\pm$ 4 & -0.120 & 2.32 & 512\\ 
 \hline
\end{tabular}
\caption{Summary of characterization results: Out-of-plane (OOP) saturation field, maximum Verdet constant (within the wavelength range of 520--\SI{530}{\nano\meter}), maximum Faraday rotation, and saturation magnetization extracted from VSM measurements.}
\label{table:2}
\end{table}

\end{comment}

\begin{table}[h]
\centering
\caption{Summary of \changed{magnetic} characterization results: Out-of-plane (OOP) saturation field \changed{of magneto-optical hysteresis}, maximum Verdet constant (at \SIrange{520}{530}{\nano\metre}), maximum Faraday rotation, and saturation magnetization extracted from \changed{in-plane} VSM measurements.}
\label{table:2}
\begin{tabular}{ |c|c|c|c|c| } 
 \hline
 Sample & OOP sat. field & Max Verdet const. & Max rotation & $\mu_0 M_\mathrm{s}$ \\ 
 & (\si{\milli\tesla}) & ($10^{8}$~\si{\degree\per\metre\per\tesla}) & ($10^{7}$~\si{\degree\per\metre}) & (\si{\milli\tesla})\\
 \hline
 1 & \num{150+-54} & \changed{\num{-1.09+-0.05}} & \changed{1.7} & \changed{\num{69+-8}}\\ 
 \hline
 2 & \num{206+-5} & \changed{\num{-0.48+-0.02}} &  \changed{1.4} & \changed{\num{95+-9}}\\ 
 \hline
 3 & \num{234+-4} & \changed{\num{-1.03+-0.04}} & \changed{2.4} & \changed{\num{217+-36}}\\ 
 \hline
 4 & \num{347+-7} & \changed{\num{-0.52+-0.02}} & \changed{1.8} & \changed{\num{157+-13}}\\ 
 \hline
 5 & \num{229+-9} & \changed{\num{-0.99+-0.04}} & \changed{2.5} & \changed{\num{189+-16}}\\ 
 \hline
 6 & \num{304+-13} & \changed{\num{-0.74+-0.07}} & \changed{2.5} & \changed{\num{174+-23}}\\ 
 \hline
 7 & \num{349+-7} & \changed{\num{-0.60+-0.03}} & \changed{2.4} & \changed{\num{187+-17}}\\ 
 \hline
 8 & \num{180+-4} & \changed{\num{-1.37+-0.06}} & \changed{2.4} & \changed{\num{164+-15}}\\ 
 \hline
\end{tabular}

\end{table}

The slopes of the Faraday rotation curve encode the magneto-optical sensitivity of the LuBiIG films. To quantify this sensitivity under the low magnetic fields that are essential for magnetic-flux imaging in low-temperature superconductors, the electromagnet and samples were first demagnetized using a decreasing alternating magnetic field and the measurements were repeated over a reduced magnetic field range well below saturation. Over this smaller magnetic field range, the magneto-optical response is linear as shown in Fig.~\ref{results}(b). As the polarization rotation $\theta$ is a linear function of the magnetic flux density $B$, and it presents almost no hysteresis, the Verdet constant $V$ is then calculated $V=\frac{1}{t} \frac{\partial \theta}{\partial B}$ where $t$ is the thickness of the film reported in Table \ref{table:1}. The values of the Verdet constants of all LuBiIG films are plotted as a function of wavelength in Fig.~\ref{results}(c), showing that the magneto-optical sensitivity peaks sharply at \SIrange{520}{530}{\nano\metre}. \changed{This maximum corresponds to electronic transitions involving hybridized Bi~6$p$--O~2$p$--Fe~3$d$ states~\cite{Shinagawa1998,Hansen1985a,Postava2000}. The Faraday effect in iron garnets originates from electric-dipole transitions of the Fe--O orbitals. The large ionic radius of Bi distorts the local oxygen coordination, while its strong spin--orbit coupling lifts the degeneracy of the excited states, splitting the transition into oppositely circularly polarized components and thereby producing a large off-diagonal permittivity and a strong dispersion of the magneto-optical response~\cite{Deb2019}. Bi-substituted iron garnets exhibit two such transitions: a higher-energy one derived from the octahedral Fe sublattice alone, and a lower-energy one to which both Fe sublattices contribute~\cite{Deb2019}. Both are resolved as steps in our transmittance spectra (Fig.~S2 of the Supplementary Material), at approximately \SI{450}{\nano\metre} and \SI{550}{\nano\metre} respectively, confirming their presence in our films. Our magneto-optical spectrometer is limited to wavelengths above \SI{500}{\nano\metre}, so only the lower-energy transition is accessible to the Faraday measurement; it is also the transition relevant to the intended operating wavelength.} As a result, employing standard solid-state lasers with the common wavelength of \SI{532}{\nano\metre} would yield near-optimal sensitivity for magneto-optical detection.

%This maximum corresponds to electronic transitions involving hybridized Bi~6$p$--O 2$p$--Fe 3$d$ states, which enhance spin-orbit coupling and produce a strong dispersion in the magneto-optical response~\cite{Hansen1985a,Shinagawa1998,Postava2000}. As a result, employing standard solid-state lasers with the common wavelength of \SI{532}{\nano\metre} would yield near-optimal sensitivity for magneto-optical detection.

%The Verdet constant $V$ is then calculated $V=\frac{1}{t} \frac{\partial \theta}{\partial B}$ where $t$ is the thickness of the film reported in table \ref{table:1}. The absolute value of the Verdet constant is plotted as a function of wavelength in Fig.~\ref{results}(b). The Verdet constant is negative, indicating a paramagnetic-type response consistent with previous reports for bismuth-substituted iron garnets~\cite{Murai2013}. The magneto-optical sensitivity peaks sharply in the range of 520--\SI{530}{\nano\metre}, decreasing rapidly at longer wavelengths. This maximum corresponds to electronic transitions involving hybridized Bi~6$p$--O~2$p$--Fe~3$d$ states, which enhance spin-orbit coupling and produce a strong dispersion in the magneto-optical response~\cite{Hansen1985a,Shinagawa1998,Postava2000}. As a result, employing standard solid-state lasers with the common wavelength of \SI{532}{\nano\metre} would yield near-optimal sensitivity for magneto-optical detection.

Table~\ref{table:2} summarizes the peak Verdet constants for all samples, which significantly exceed the values typically reported for conventional iron garnets~\cite{Carothers2022} and rank among the highest achieved for LuBiIG thin films~\cite{Veshchunov2016,Thakur2023}. The largest Verdet constants are observed for samples~1 and~8 (\changed{\SI{-1.09e8}{\degree\per\metre\per\tesla}} and  \changed{\SI{-1.37e8}{\degree\per\metre\per\tesla}}, respectively), \changed{which also exhibit the lowest out-of-plane (OOP) magnetic saturation fields.} 

\changed{Since the OOP loop is nearly linear between saturation fields, the Verdet constant of these films is set by the ratio of the maximum Faraday rotation to the OOP saturation field. The Verdet constant can therefore be improved either by increasing the rotation or by reducing the saturation field. The OOP saturation field has contributions from dipole-field-induced shape anisotropy, magneto-crystalline anisotropy and strain-induced anisotropy. Were shape anisotropy the only contribution, the OOP saturation field would be comparable to $\mu_0 M_\mathrm{s}$, i.e.\ around \SI{170}{\milli\tesla}. The measured values of \SIrange{150}{350}{\milli\tesla} therefore indicate the presence of additional anisotropy terms. The XRD data show a compressive in-plane strain which, given the negative magnetostriction of the film\cite{Das2023}, adds to the OOP saturation field, with the more coherently strained films tending to show larger saturation fields. The maximum Faraday rotation, by contrast, depends
mainly on the local bond order and is much less sensitive to crystalline quality. The net effect is that the largest Verdet constants occur in the films of reduced crystallinity, samples~1 and~8, where the spread seen in the RSM of Fig.~\ref{fig:xrd}(b) is consistent with strain relaxation and hence a lower OOP saturation field. This also indicates a route to further improvement: growing on a substrate with a larger lattice parameter, such as gadolinium scandium gallium garnet\cite{Fratello1987}, would impose a tensile
in-plane strain that lowers the OOP saturation field while preserving high crystalline quality, potentially yielding an order of magnitude improvement in the Verdet constant.}

\begin{comment}

\changed{For comparison, the \hl{commercially-available? Why is this a comparator?} Verdet constant of TGG (Terbium Gallium Garnet) at room temperature was reported\cite{Slezak2016} to be $1.4\times 10^{4}  \rm\,^\circ m^{-1} T^{\,-1}$, and other rare-earth substituted garnets not containing bismuth have shown\cite{Wu2023} constants as high as $3.4\times 10^{4} \rm\,^\circ m^{-1} T^{\,-1}$. Among lutetium- and bismuth-substituted garnets, the Verdet constant of Lu$_{1.2}$Bi$_{1.8}$IG was measured\cite{Hansen1985} to be $-2.5\times 10^{7}  \rm\,^\circ m^{-1} T^{\,-1}$ at room temperature and $633\,\rm nm$, and Ref.~\citenum{Thakur2023} makes use of a LuBiGaFeO material with a Verdet constant of $-1.6\times 10^{8}  \rm\,^\circ m^{-1} T^{\,-1}$ at $550\,\rm nm$. Therefore, the films on which this work reports are more magneto-optically active by a few orders of magnitude than typical iron garnets and in fact stand among the most magneto-optically active materials ever reported.}\\ 
\end{comment}

\changed{For comparison, terbium gallium garnet (TGG) is the benchmark material of commercial Faraday isolators and rotators. As a paramagnet, its Verdet constant at room temperature is only \SI{1.4e4}{\degree\per\metre\per\tesla}\cite{Slezak2016}, and other rare-earth garnets not containing bismuth reach at most \SI{3.4e4}{\degree\per\metre\per\tesla}\cite{Wu2023}. The ferrimagnetic, bismuth-substituted garnets are three to four orders of magnitude more active: for Lu$_{1.2}$Bi$_{1.8}$IG a value of \SI{-2.5e7}{\degree\per\metre\per\tesla} was measured at room temperature and \SI{633}{\nano\metre}\cite{Hansen1985}, while a LuBiGaFeO garnet reaches \SI{-1.6e8}{\degree\per\metre\per\tesla} at \SI{550}{\nano\metre}\cite{Thakur2023}. The films reported here, at \SI{-1.37e8}{\degree\per\metre\per\tesla}, thus exceed previously reported LuBiIG films by a factor of five and approach the largest Faraday response documented for any material.}

\changed{While the maximum rotation values we report are large, they are consistent with previous literature. A linear increase of Faraday rotation with Bi content has previously been reported \cite{Hansen1985a,Helseth2001}. Ref.~\citenum{Helseth2001} predicts a maximum rotation of \SI{1.1e7}{\degree\per\metre} at \SI{539}{\nano\metre} based on their LPE-grown films. That we are able to obtain values up to \SI{2.5e7}{\degree\per\metre} is due to the use of PLD growth, which avoids the reduction in Faraday rotation caused by the incorporation of PbO from the LPE flux \cite{Johansen1976}.}

To give an example of these films' potential applicability, we consider a vortex in a superconducting sample with a penetration depth of about \SI{100}{\nano\metre}, typical of thin films. At a distance of about \SI{200}{\nano\metre} of the superconductor's surface where an indicator film could reasonably be positioned using flip-chip stacking, a magnetic field anomaly of some \SI{0.1}{\milli\tesla} can be expected due to the vortex. Using a film with the parameters of sample 8 of this work together with an illumination wavelength  \SI{532}{\nano\metre} and a power of \SI{1}{\micro\watt} which is expected to conserve the superconducting properties of the sample \cite{Veshchunov2016}, we estimate that a cross-polarizer setup would yield a signature of about \SI{1}{\pico\watt} of optical power anomaly. This power is well above the electronic detection noise of most modern photodiodes. Assuming that the cross-polarizer setup leaks \SI{1}{\percent} of the \SI{1}{\micro\watt} reflected intensity, the shot-noise from that parasitic signal would be of the order of \SI{50}{\femto\watt}$/\sqrt{\mathrm{Hz}}\ll$~\SI{1}{\pico\watt}, setting a sensitivity estimate \changed{easily} sufficient to detect the signature from a single vortex, %although this estimation assumes that the magneto-optical coupling retains its value 
\changed{even if the magneto-optical sensitivity does not increase at lower temperatures, as is generally the case for similar materials \cite{Deb2019, Hansen1985}}.\\

%\section{Conclusion}
%The large saturation fields will in principle allow for linear and reliable (non-hysteretic) magneto-optical detection over a broad range of magnetic field values.

\changed{In summary, we reported on (i)~the growth of crack-free, single-phase LuBiIG films with thicknesses of \SIrange{85}{220}{\nano\metre} grown on sGGG(111) using pulsed laser deposition, identifying a 2D heteroepitaxial growth window; (ii)~XRD/RSM analyses of these films confirming fully coherent, pseudomorphically-strained growth for the best films; (iii)~EDS analyses verifying that the 2D-growth films match the nominal $\mathrm{Bi}_{2.1}\mathrm{Lu}_{0.9}\mathrm{Fe}_5\mathrm{O}_{12}$ stoichiometry within the resolution of the technique; and (iv)~the demonstration of a maximum Verdet constant of \SI{-1.37e8}{\degree\per\metre\per\tesla} peaking at \SIrange{520}{530}{\nano\metre}.} \changed{These parameters establish optimized PLD-grown LuBiIG films as a benchmark} for next-generation magneto-optical imaging and field-sensing technologies.

\section*{Supplementary Material}
\changed{The supplementary material details the methodology and results of the characterization of samples by RHEED, EDS, transmittance measurement, AFM, XRD, and reports on the confirmation of the magnitude of Faraday rotations measured in the article by measuring them independently in a rotating-analyzer polarimeter.}

\section*{Acknowledgements}
\changed{We thank Junyoung Hyun and Sreeveni Das} for their help and guidance on various characterization techniques.
We acknowledge the facilities and technical support of
Otaniemi research infrastructure for Micro and Nanotechnolo-
gies (OtaNano). We also acknowledge the financial support of the Finnish Ministry of Education and Culture through the Quantum Doctoral Education Pilot Program (QDOC VN/3137/2024-OKM-4) and the Research Council of Finland through the Finnish Quantum Flagship project (project number 358877, Aalto University). L. M. d. L. acknowledges funding from the Research Council of Finland (project number 338565).

\section*{Conflicts of interest}
The authors have no conflicts of interest to disclose.

\section*{Author contribution}
\textbf{Megan H.\ Dransfield}: Formal analysis (equal), Investigation (equal), Writing - original draft (lead), Writing - review and editing (equal).
\textbf{Lukáš Flajšman}: Conceptualization (supporting), Formal analysis (equal), Investigation (equal), Writing - review and editing (equal), Funding acquisition (supporting), Resources (lead).
\textbf{Matthijs H.\ J.\ de Jong}: Writing - review and editing (equal), Investigation (equal), Supervision (equal).
\changed{\textbf{Lars Peeters}: Investigation (equal).}
\changed{\textbf{Rhodri Mansell}: Writing - review and editing (equal), Investigation (equal).}
\textbf{Sebastiaan van Dijken}: Supervision (supporting), Resources (supporting), Writing - review and editing (equal).
\textbf{Laure Mercier de Lépinay}: Conceptualization (lead), Funding acquisition (lead), Supervision (equal), Writing - review and editing (equal).

\section*{Data Availability Statement}
The data that support the findings of this study are available from the corresponding author upon reasonable request.

\section*{References}

\bibliographystyle{apsrev4-2}
\bibliography{lubigpaper}

\end{document}